\documentclass{INTERSPEECH2023}

\interspeechcameraready

\usepackage{amsmath}
\usepackage{algorithm}
\makeatletter
\newcommand{\algrule}[1][.2pt]{\par\vskip.5\baselineskip\hrule height #1\par\vskip.5\baselineskip}
\makeatother
\usepackage{algpseudocode}

\title{Iteratively Improving Speech Recognition and Voice Conversion}
\name{Mayank Kumar Singh$^1$, Naoya Takahashi$^2$, Onoe Naoyuki$^1$}
\address{
  $^1$Sony Research India, India, $^2$Sony Group Corporation, Japan}
\email{\{mayank.a.singh, naoya.takahashi, naoyuki.onoe\}@sony.com}

\begin{document}

\maketitle
 
\begin{abstract}
  Many existing works on voice conversion (VC) tasks use automatic speech recognition (ASR) models for ensuring linguistic consistency between source and converted samples. However, for the low-data resource domains, training a high-quality ASR remains to be a challenging task. In this work, we propose a novel iterative way of improving both the ASR and VC models. We first train an ASR model which is used to ensure content preservation while training a VC model. In the next iteration, the VC model is used as a data augmentation method to further fine-tune the ASR model and generalize it to diverse speakers.
  By iteratively leveraging the improved ASR model to train VC model and vice-versa, we experimentally show improvement in both the models.
  Our proposed framework outperforms the ASR and one-shot VC baseline models on English singing and Hindi speech domains in subjective and objective evaluations in low-data resource settings.
  
  
  
\end{abstract}

\section{Introduction}


Speech processing technologies such as voice conversion (VC) and automatic speech recognition (ASR) have dramatically improved in the past decade owing to the advancements in deep learning technologies. However, the task of training these models remains challenging on low resource domains as they suffer from over-fitting and do not generalize well for practical applications. As VC models often rely on ASR model for extracting content features or imposing content consistency loss, \cite{ucdsvc, starganv2vc, rosvc}, degradation of ASR directly affects the quality of VC models. On the other hand, to improve the generalization capability of ASR model, a variety of data augmentation techniques have been proposed with voice conversion being one of them \cite{google_data_aug_vc}.

This creates a causality dilemma, wherein poor quality of ASR model affects VC model training which in turn leads to low quality data augmentation for training ASR models. Conversely, improving the ASR model should lead to better VC models, which should produce better data augmentation samples for improving the ASR models. Motivated from this, in this work we propose to iteratively improve the ASR model by using the VC model as a data augmentation method for training the ASR and simultaneously improve the VC model by using the ASR model for linguistic content preservation. 

We validate the proposed framework on two low-resource domains, namely Hindi speech voice and English singing domains, and show that our proposed iterative framework improves the WER of ASR model as well as improves the subjective and objective metrics of the VC model. We chose English singing voice as it has low amount of public resources for the task of VC as well as low amount of manual annotations for training an ASR. Acapella singing voice without annotations is also scarce, making the training of unsupervised ASR models unfeasible. For showing the applicability of our approach across languages different from English, we chose the Hindi speech domain in low resource settings.
\\
\\
The contributions of our work are summarized as follows:
\begin{enumerate}
\item We propose an iterative framework to improve ASR models by using VC models for data augmentation and VC models by using ASR models as speech consistency loss.
\item We evaluate the proposed framework on Hindi speech domain under low data resource settings and successfully show its superiority over the baseline models for both ASR and VC.  
\item We evaluate the proposed framework for the one-shot singing voice conversion and singing voice recognition task using NUS-48E and NHSS datasets and show that the proposed framework significantly improves the word error rate of the ASR model and quality of samples generated using the VC model compared to models trained without the proposed framework.
\end{enumerate}

The audio samples are available at our website\footnote{\url{https://demosamplesites.github.io/IterativeASR\_VC/}.}

\section{Related Work}

The problem of converting the speaker identity of a voice while preserving the linguistic content, such as speech voice conversion (converting the speaker identity of speech signal) \cite{starganv2vc, autovc, adainvc, againvc, vqvc+}, singing voice conversion (converting the speaker identity of a singing signal) \cite {hvqvc, ucdsvc, rosvc, bytedance} and emotional voice conversion \cite{evcup} (converting the emotion of expressive speech signals) has been actively studied for decades.
Recent advances in deep learning have been inspired from techniques like auto-encoder \cite{autovc, adainvc, hvqvc, bytedance} and generative adversarial networks (GAN) \cite{starganv2vc, rosvc}. Recently, ASRs have been used to improve the linguistic content preservation of VC models, \cite{ucdsvc, starganv2vc, rosvc}. These techniques have dramatically improved VC models, yet the performance drops significantly when we consider the challenging and realistic scenario of working on low data resource domains.

ASR performance in the supervised case has seen rapid improvements in the last few years due to advances in the model architecture \cite{ASR1, ASR2, ASR3}, training methods \cite{wav2vec2} and increasing data size \cite{ls_asr_dataset}. Recently, unsupervised ASR has gained popularity \cite{wav2vecU} and shown remarkable efficacy in utilizing large sets of unlabelled data. Yet, the accuracy of the ASR models drop significantly when we try to generalize them across domains, for e.g. from speech to singing, across languages, or train them on low data resource domains. It becomes more important to address the limitations of current ASR models because of their usage in domains having wide variety of applications like voice conversion \cite{ucdsvc, rosvc, google_data_aug_vc}, source separation \cite{voicesep} and text to speech generation \cite{tts}.

\section{VC Data Augmentation in ASR training}

Data augmentation has been effectively used for training ASR models by modifying the pitch \cite{data_aug_asr}, applying spectral augmentation and speed perturbation \cite{speed_perturb}. Although these augmentation strategies have been robustly shown to improve ASR models, recently, voice conversion has been shown to outperform these methods \cite{vc_in_asr1, google_data_aug_vc}. The effectiveness of using VC to augment the dataset in very low resource domains and superiority of using VC over spectral augmentation is established in \cite{vc_in_asr1}.
However, they do not make use of an ASR model during the training of the VC model, hence their VC model cannot be shown to leverage the improved ASR model which is trained using VC data augmentation. In \cite{google_data_aug_vc}, it is shown that even in the case of high data resource domain, VC is effective as a data augmentation method for generalizing the ASR model to unseen speakers. They replace the encoder of the VC model with a pre-trained ASR encoder and show the effect of using such an encoder on the quality and speaker identity conversion. This motivated us to adopt an iterative approach of using the VC as a data augmentation tool for improving the ASR and using the ASR to improve the VC model. Different from \cite{google_data_aug_vc}, which uses the ASR encoder to disentangle the speaker identity and the linguistic content, we use the ASR encoder as an auxiliary loss to preserve the linguistic content. We further detail this in Section \ref{sec:speech_consis_loss}.

\section{Speech Consistency Loss}
\label{sec:speech_consis_loss}

\begin{figure}[t]
  \centering
  \includegraphics[width=\linewidth]{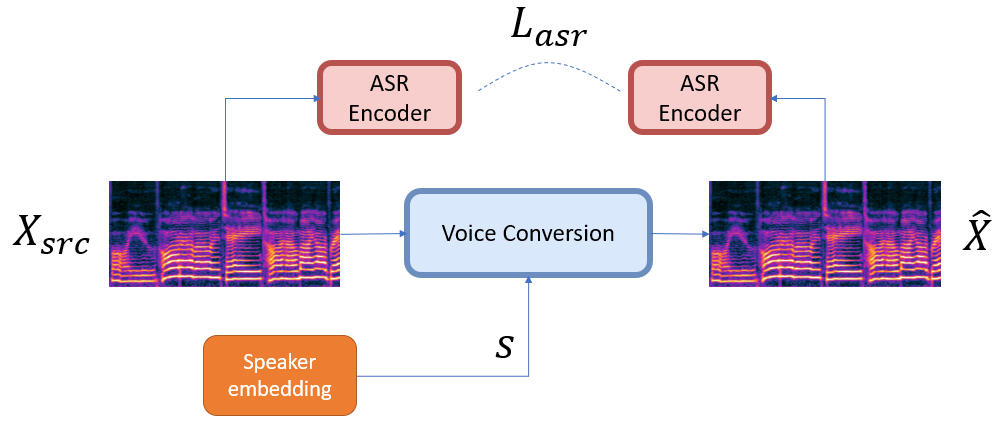}
  \caption{VC Model training with Speech Consistency Loss.}
  \label{fig:vc_1_step}
  \vspace{-5mm}
\end{figure}

The effectiveness of using a pre-trained ASR model for preserving the linguistic content during voice conversion has been shown in \cite{ucdsvc, starganv2vc, google_data_aug_vc}. 
One way of using ASR features to preserve the linguistic features is by applying the speech consistency loss \cite{starganv2vc, ucdsvc}. In \cite{starganv2vc}, the speech consistency loss is defined as

\begin{equation}
    \mathcal{L}_{asr}(\theta)=\mathbb{E}_{X,s}\Big[ ||h_{asr}(X) - h_{asr}(V_{\theta}(X,s))||_{1} \Big]
\end{equation}
where $X$ denotes mel-spectrogram of the source, $h_{asr}(\cdot)$ denotes the encoder features of a differentiable pre-trained ASR model, $s$ is the speaker embedding of the target speaker, and $V$ is the voice conversion model which is conditioned upon $X$ and $s$. $\theta$ denotes the VC model parameters.
It is assumed that the output of the encoder of an ASR model is independent of speaker identity and pitch and contains only the linguistic content. Thus minimizing the $L_{asr}$ loss emphasises the preservation of the linguistic content after voice conversion and does not hinder speaker identity conversion nor pitch conversion.

The voice conversion training process using an ASR model for preserving the linguistic content is shown in Figure~\ref{fig:vc_1_step}.

\begin{figure}[t]
  \centering
  \includegraphics[scale=0.45]{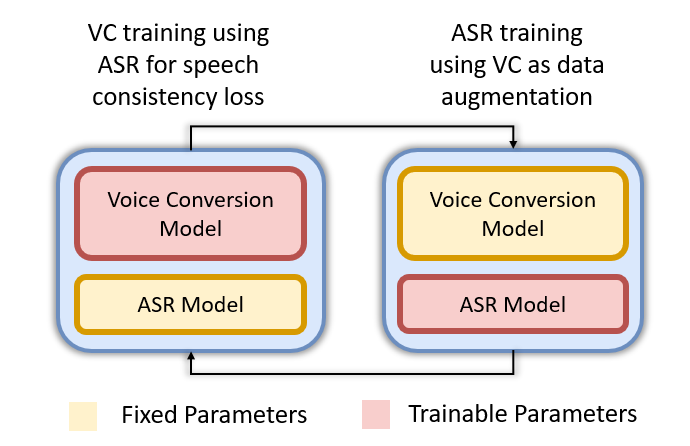}
  \caption{Iterative training for VC and ASR}
  \label{fig:iterative_fig}
  \vspace{-5mm}
\end{figure}

\section{Iterative Training Framework}
\begin{algorithm}
\caption{Iterating training of ASR and VC}\label{psuedocode}
\begin{algorithmic}[1]
\Require{Training data set $\tau$, ASR fine-tuning routine $F(A_{i}, Dataset)$, VC training routine $T(Dataset, A_{i})$}
\algrule
\State Train $A_{0}$ using standard ASR training
\State Train $V_{0}$ by utilizing $A_{0}$ for minimizing $L_{asr}$
\State i = 0
\While{not done}
\State $\hat{\tau}_{i+1}$ = $V_{i}(\tau )$ \Comment{Apply VC data augmentation}
\State $A_{i+1} = F(A_{i}, \{\tau, \hat{\tau}_{i+1}\})$ \Comment{Fine-tune ASR Model}
\State $V_{i+1} = T(\tau , A_{i+1})$ \Comment{Train VC Model}
\State $i \gets i + 1$
\EndWhile
\end{algorithmic}
\end{algorithm}
The proposed iterative training framework is illustrated in Figure~\ref{fig:iterative_fig} and its pseudo code is shown in Algorithm \ref{psuedocode}. In the first step of the iterative training, an ASR model is trained on the low resource domain dataset, $\tau$, which we call as $A_{0}$. Next, we train a VC model which uses $A_{0}$ for minimizing $L_{asr}$. We call this VC model as $V_{0}$. In the next step, we use $V_0$ to apply data augmentation to $\tau$ and get $\hat{\tau}_{1}$. For applying the VC data augmentation, we sample both the source and reference from the train dataset and provide them to $V_0$ to artificially increase the diversity of the training samples in terms of speaker style. We further fine-tune $A_{0}$ on $\tau$ and $\hat{\tau}_{1}$ and refer to the fine-tuned model as $A_{1}$. Using $A_{1}$, for minimizing $L_{asr}$, we again train a VC model which we denote as $V_{1}$. This process of iterative refinement of ASR and VC models is repeated until the WER of the ASR model converges on the validation set.

The motivation of our approach is as follows. For training a VC model, using an ASR model which is able to extract better linguistic features and generates speaker independent encoder features will lead to better content preservation and clarity. Similarly, for training an ASR model, using a VC model which preserves more linguistic content and produces less artefacts will generate more realistic samples for data augmentation. By iteratively training the ASR and VC model we leverage the improved models for training the next iteration and thus achieve better performance on the objective and subjective metrics.

Although in this work we use speech consistency loss to leverage the ASR model for improving the linguistic preservation capacity of the VC models, another approach of using the ASR models is by providing the ASR features extracted from $X_{src}$ as an input to the VC model. Such an approach is used in \cite{ucdsvc}, which is an auto-encoder based VC model that takes the pitch, loudness and ASR features as input. Although this approach has the advantage of utilizing non-differentiable ASR models for training the VC model, it strongly relies on the ASR features for extracting the linguistic content that can lead to poor generalization of the VC model to out-of-domain samples. 

\section{Experiments}

To evaluate the proposed framework in a realistic scenario of low resource domains we chose Hindi speech and English singing domains for the task of voice conversion and automatic speech recognition.

\subsection{Datasets}

For the Hindi speech dataset, we use an internal data consisting of nine speakers with a total duration of three hours. Out of the nine speakers, five are male and four are female. The speakers were given non-overlapping transcripts in the Hindi language and were asked to speak them in a neutral tone. The recordings were conducted in a studio setup. As the transcripts are in Hindi, we use a transliteration package \cite{hin_trans} to convert the Devanagri characters to Latin characters. For training, we use a subset of seven speakers and use two speakers for evaluation. We train an ASR model for WER objective evaluation of the Hindi speech VC models. For this, we use the publicly available MUCS2021 dataset \cite{hindiasreval} having 95.05 hours of training audio and 5.55 hours of testing audio.

For the English singing dataset, we use the publicly available NUS-48E~\cite{nus} and NHSS~\cite{nhss} datasets. The NUS-48E dataset consists of twelve singers with a total of two hours of singing data and one hour of speech data. The NHSS dataset consists of ten singers with a total of 4.75 hours of singing data and 2.25 hours of speech data. For the experiments we use only the singing data from NUS-48E and NHSS datasets. For training, we use a subset of ten singers from NUS-48E dataset and eight singers from NHSS dataset which has a duration of four hours and use the remaining speakers for evaluation.

For both the Hindi speech and English singing domain task, the ASR model is trained on a combination of the target domain and LibriSpeech dataset, which is a large English dataset consisting of 960 hours of data \cite{librispeech} and is publicly available. The LibriSpeech corpus is incorporated during training to improve generalization of the ASR model.

For the VC and ASR training, all the data is first re-sampled to 24kHz and an 80 bin mel-spectrogram representation is calculated as in \cite{starganv2vc}.

\begin{table}[t]
   \caption{Hindi ASR WER across iterations}
  \label{tab:tab1}
  \centering
  \begin{tabular}{c | c c c c}
    \toprule
    Iter & $0$ & $1$ & $2$ & $3$ \\
    \midrule
    WER $\downarrow$ &\hspace{-1.5mm} $27.6 \pm 0.13$\hspace{-2.5mm} & $26.2 \pm 0.17$\hspace{-2.5mm} & $25.8 \pm 0.16$\hspace{-2.5mm} & $\textbf{25.7} \pm \textbf{0.15}$\hspace{-1mm}\\
    \bottomrule
  \end{tabular}
  \vspace{-5mm}
\end{table}

\subsection{Model Architecture}

For the ASR model, we choose the ESPNet framework \cite{espnet} having a hybrid CTC-Attention model \cite{espnet_hybrid}. For $h_{asr}$, we use the output of the encoder with 512 dimensions. As the ASR model needs to extract linguistic features from the VC model's converted samples, we modify the mel-spectrogram extraction hyper-parameters to match those of the VC model. 

For the VC model we adopted \cite{rosvc}, which is a recently proposed modification of \cite{starganv2vc}, a GAN-based network. A brief overview of the VC model architecture is provided in Figure \ref{fig:vc_1_step}. The implementation details are the same as in \cite{rosvc}.


\subsection{Training Details}

\textbf{ASR}\hspace{2mm} For training $A_{0}$ we first train an ASR on the LibriSpeech dataset, which we call as $A_{base}$ and then fine-tune it on the low resource domain. The $A_{base}$ model was trained for 35 epochs with a batch size having cumulative duration of 30 minutes. We use Adam \cite{adam} optimizer with a learning rate of 0.0025 with a warm-up scheduler of 40k iterations. For each iteration in the iterative training, we fine-tune the ASR model for a total of 100k iterations with a learning rate of 0.00025. During fine-tuning, we keep the distribution of the LibriSpeech dataset to that of the target dataset as 1:3. The training details remain the same across iterations.\\
\textbf{VC}\hspace{2mm} We train the VC model with the same hyper-parameters as in \cite{rosvc}, except for the weight for the $L_{asr}$ which we set as 100 for all the experiments. We use the hierarchical prior grad vocoder, proposed in \cite{hpg}, to synthesize waveform from mel-spectrogram.
All the experiments are conducted using 2 A-100 Nvidia GPUs having runtime of ~24 hour.

\subsection{Evaluations}

For showcasing the effectiveness of our proposed framework, we consider $A_{0}$ as a baseline model for the ASR evaluations and $V_{0}$ as a baseline model for the VC evaluations. We evaluate our ASR models using the WER metric. For evaluating the VC models, we use both objective as well as subjective metrics.

\begin{table}[t]
  \caption{Hindi VC Results.}
  \label{tab:tab2}
  \centering
  \begin{tabular}{c | c c c}
    \toprule
    Iterations & WER $\downarrow$ & pMOS $\uparrow$ & Identity $\uparrow$ \\
    \midrule
    0 & $41.93 \pm 0.07$ & $3.79 \pm 0.01$ & $0.7410 \pm 0.0007$ \\
    1 & $38.87 \pm 0.07$ & $3.89 \pm 0.01$ & $0.7388 \pm 0.0007$ \\
    2 & $38.85 \pm 0.07$ & $3.91 \pm 0.01$ & $0.7282 \pm 0.0007$ \\
    3 & $\textbf{38.35} \pm \textbf{0.07}$ & $\textbf{3.94} \pm \textbf{0.01}$ & $\textbf{0.7498} \pm \textbf{0.0007}$ \\
    \bottomrule
  \end{tabular}
  \vspace{-5mm}
\end{table}

\textbf{Objective metric for VC.}\hspace{2mm} We use NISQA-TTS \cite{nisqa} for predicted mean opinion score(pMOS). For evaluating the speaker conversion of the VC models, we use the Identity metric, which is the cosine similarity of d-vectors extracted using Resemblyzer \cite{resemblyzer}, a speaker verification model. For calculating how well the VC model preserves linguistic content, we calculate the WER using an ASR model which is different from the ASR model used for training the VC model.
For the Hindi speech domain, we train an ASR model on the MUCS 2021 dataset which is publicly available.
For English singing domain, we use an ASR model trained using iterative framework and having different distribution of speech and song data during training.
For the objective and subjective test, we use 346 and 420 samples for the Hindi speech and English singing domain, respectively. The samples have a duration between 2 and 12 seconds.

\textbf{Subjective metric for VC.}\hspace{2mm} We ask 17 audio engineers for the Hindi speech and 23 audio engineers for the English singing domain domain evaluation. The evaluators have no known hearing impairment. For the Hindi speech AB and SMOS test, the evaluators evaluate a total of 170 samples per model. For the English singing AB and SMOS test, the evaluators evaluate a total of 230 utterances per model. 

\section{Results}

We present the results of the objective and subjective evaluation along with their 95\% confidence interval.

\subsection{Objective Evaluation}
\subsubsection{Hindi speech voice dataset}

\textbf{ASR}\hspace{2mm} 
The WER of the Hindi ASR model is reported in Table \ref{tab:tab3}. 
The Hindi ASR model achieves the best WER of 25.8 for iteration 3 with a relative improvement over the ASR model trained without iterative training of $6.85\%$.\\
\textbf{VC}\hspace{2mm} Objective test results for Hindi speech VC model is reported in Table \ref{tab:tab4} which shows a decrease in WER and increase in pMOS across iterations.

\begin{table}[t]
  \caption{Singing ASR WER across iterations. Note that we stop training the ASR after the WER converges.}
  \label{tab:tab1}
  \centering
  
  \begin{tabular}{c | c c c c}
    \toprule
    Iter & $0$ & $1$ & $2$ & $3$ \\
    \midrule
    WER $\downarrow$ & $6.7 \pm 0.07$ & $6.0 \pm 0.1$ & $5.0 \pm 0.1$ & $\textbf{4.9} \pm \textbf{0.1}$ \\
    \bottomrule
  \end{tabular}
\end{table}

\begin{table}[t]
  \caption{Singing VC Results.}
  \label{tab:tab3}
  \centering
  \begin{tabular}{c | c c c}
    \toprule
    Iterations & WER $\downarrow$ & pMOS $\uparrow$ & Identity $\uparrow$ \\
    \midrule
    0 & $20.83 \pm 0.22$ & $3.11 \pm 0.01$ & $0.7701 \pm 0.001$ \\
    1 & $19.12 \pm 0.21$ & $3.11 \pm 0.01$ & $0.7794 \pm 0.001$ \\
    2 & $\textbf{17.55} \pm \textbf{0.21}$ & $\textbf{3.38} \pm \textbf{0.01}$ & $0.7807 \pm 0.001$ \\
    3 & $17.56 \pm 0.20$ & $3.36 \pm 0.01$ & $\textbf{0.7808} \pm \textbf{0.001}$ \\
    \bottomrule
  \end{tabular}
  \vspace{-5mm}
\end{table}


\subsubsection{English singing voice dataset}

\textbf{ASR}\hspace{2mm} We evaluate our proposed framework on the English singing voice domain and report the WER in Table \ref{tab:tab1}.
Similar to the Hindi speech domain we observe that after 3 iterations the WER of the ASR model converges. We validate our proposed framework and show that the WER converges to $4.9$ after 3 iterations with a relative improvement over the ASR model trained without our proposed framework of $26.8\%$.
\\
\textbf{VC}\hspace{2mm} For the VC evaluation of our proposed framework, we report the results in Table \ref{tab:tab2}. 
We validate our proposed iterative training framework and show a relative decrease in WER of 12.63\% and relative increase in the pMOS of 8.83\%. Across iterations, the Identity metric remains nearly unchanged.



\subsection{Subjective Evaluation}

For the Hindi speech domain, we compare the $V_3$ model with the baseline $V_0$ model based on the results of the objective evaluation. Similarly, for the English singing domain, we compare the $V_2$ model with the baseline $V_0$ model. For subjective evaluation we consider AB quality test and Similarity MOS (SMOS). The AB quality test is constructed to compare the quality of the samples generated using the proposed model with the samples generated using the baseline model. During evaluation, we present one audio sample each from the baseline and proposed model to the evaluator and ask them to rate their preference in terms of quality. The two audio samples are generated using the same source and reference samples. The SMOS test is constructed to evaluate the speaker similarity of the converted sample to that of the reference sample. The evaluators are presented the reference sample and the voice converted sample and asked to rate the similarity of the voices between the samples on a scale of 1-5. We also evaluate the SMOS for oracle samples in which the reference and converted are real samples from the same speaker.
The results of the subjective test for the VC models are summarized in Table \ref{tab:tab5}. As we can observe, the proposed iterative training outperforms the baseline models on AB quality test with a preference rate of 0.66 to 0.34 on Hindi speech VC and 0.61 to 0.39 on English singing VC while achieving better SMOS.

\subsection{Observations}

By using the improved ASR for training the VC models we expected a decrease in the WER of the samples after conversion which is validated in Table \ref{tab:tab2} and \ref{tab:tab4}. However, we also observe that the quality of the converted samples improve via the proposed framework due to decrease in artefacts and clearer pronunciation of words. This further demonstrates that with decrease in WER of the ASR model, the encoder features better capture the linguistic content which is essential for improving the quality of VC models. Because of the improvement in the quality of the converted samples because of less artefacts and clearer pronunciation, we expected the VC models to act as better data augmentation tools for training the ASR model which is validated from the results in \ref{tab:tab1} and \ref{tab:tab3}.



\begin{table}[t]
  \caption{Subjective Test Results for SMOS.}
  \label{tab:tab3}
  \centering
  \begin{tabular}{c | c | c c}
    \toprule
    Model & ASR Domain & SMOS $\uparrow$ \\
    \midrule
    Oracle Hindi & - & $4.68 \pm 0.14$ \\
    Hindi & Hindi Iteration 0 & $2.93 \pm 0.16$ \\
    Hindi & Hindi Iteration 3 & $\textbf{3.01} \pm \textbf{0.18}$ \\
    \midrule
    Oracle Singing & - & $4.70 \pm 0.09$ \\
    Singing & Singing Iteration 0 & $3.06 \pm 0.13$ \\
    Singing & Singing Iteration 2 & $\textbf{3.17} \pm \textbf{0.14}$ \\
    \bottomrule
  \end{tabular}
  \vspace{-3mm}
  
\end{table}

\begin{figure}[t]
    \centering
    \includegraphics[width=\linewidth]{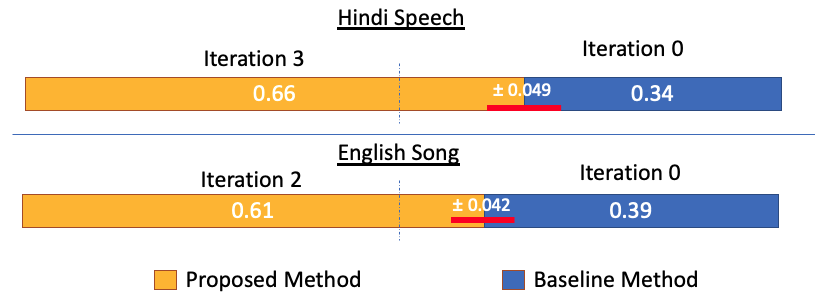}
    \caption{Subjective AB Test results for Hindi Speech Iteration 3 vs Iteration 0 and English Singing Iteration 2 vs Iteration 0}
    \label{fig:ABTest}
    \vspace{-5mm}
\end{figure}

  

\section{Conclusions}

We present a novel iterative framework for improving voice conversion models and automatic speech recognition models on low resource domains and verify its applications on the Hindi speech domain and English singing domain. We show improved speech preservation and MOS quality of the converted samples on voice conversion tasks as well as improved the word error rate on ASR tasks using this framework. 
Future work includes further improving the content preservation of the one-shot VC models so as to bring WER of the VC converted samples closer to the WER on the ground truth samples which would also lead to better MOS quality. We would also like to investigate combining the ASR and VC training in an end-to-end system.

\pagebreak
\bibliographystyle{IEEEtran}
\bibliography{main}

\end{document}